\documentclass[12pt]{article}

\title{ \vspace*{-0.3in}  \textbf{\Large  Parental Health Penalty on Adult Children's Employment: Gender Difference and Long-Term Consequence} }

   \author{\href{https://wjyecon.weebly.com/}{\textbf{Jiayi Wen}}\thanks{E-mail: wjyecon@gmail.com. School of Economics and Wang Yanan Institute of Studies in Economics (WISE), Xiamen University, Fujian, China. I acknowledge the grants from the National Natural Science Foundation of China (Grant No. 71903167) and The Fundamental Research Funds for the Central Universities (Grant No. 20720221013). All remaining errors are mine.}, \ \  \textbf{Haili Huang }}

\date{\small{Wang Yanan Institute for Studies in Economics (WISE)\\ and School of Economics, Xiamen University}  \\ \quad \\August 18, 2023 \\ \href{https://wjyecon.weebly.com/}{Here for the latest version} }

   \usepackage{natbib}
   \usepackage{amsmath}
   \usepackage{indentfirst}
   \usepackage{graphicx} 
   \usepackage{float}
   \usepackage{caption}
\usepackage[margin=1in]{geometry}
   \usepackage[justification=centering]{caption}
   \usepackage[normalem]{ulem}
   \usepackage[toc,page,title,titletoc,header]{appendix}
   \usepackage{multirow}
   \usepackage{epstopdf}

   \usepackage{booktabs}
   \usepackage{mathtools}
   \usepackage{rotating}
   \usepackage[colorlinks=true,linkcolor=blue,urlcolor=black,bookmarksopen=true]{hyperref}
   \usepackage{bookmark}
   \usepackage[titletoc]{appendix}   
   \usepackage{soul}    
   \usepackage{setspace}
   \usepackage{numprint}

\usepackage[table,xcdraw]{xcolor}   

\usepackage{array}
\usepackage{multirow}

\usepackage{tikz}
\usetikzlibrary{decorations.pathreplacing}
\tikzset{tick/.style={draw, rectangle, minimum width=0pt, minimum height=2mm, inner sep=0pt}}

\npdecimalsign{.}
\nprounddigits{3}

\hypersetup{
  colorlinks,
  citecolor=[RGB]{0,102,204},
  linkcolor=[RGB]{0,0,0},
  urlcolor=black}

\makeatletter

\newcommand{\Rmnum}[1]{\expandafter\@slowromancap\romannumeral #1@}
\makeatother

\begin{document}
\maketitle

\begin{abstract}
 This paper examines the long-term gender-specific impacts of  parental health shocks on adult children's employment in China. We build up an inter-temporal cooperative framework to analyze household work decisions in response to parental health deterioration. Then employing an event-study approach, we establish a causal link between parental health shocks and a notable decline in female employment rates.  Male employment, however, remains largely unaffected. This negative impact shows no abatement up to eight years that are observable by the sample. These findings indicate the  consequence of ``growing old before getting rich'' for developing countries.  \\


\end{abstract}

\newpage
\setstretch{1.25}
  
\section{Introduction}
Both developed and developing countries are experiencing a demographic shift towards an aging population. As evident from the seventh national census of China in 2021, individuals aged 60 and above now constitute 264 million people, representing 18.70\% of the total population, a figure that has doubled since 2000. Notably, this number surpasses the entire population of the world's fifth most populous nation. Projections indicate that by 2050, this proportion will escalate to a staggering 40\%. Within this context, health risks have emerged as a prominent concern given their upward trajectory across the life cycle.

This paper seeks to uncover the repercussions of parental health shocks on the labor supply of younger cohorts in China. In the backdrop of limited provisions of  market and public insurance, the family assumes a vital role in risk-sharing among developing countries (\cite{liu2016insuring}, \cite{fafchamps2003risk}).
Through this mechanism, health risks stemming from population aging may extend their reach beyond the elderly and impact younger generations. Intrahousehold specialization can induce an unequal distribution of the costs, particularly towards females (\cite{dercon2000sickness}). While existing studies based on developed countries find no evidence of such spillover effects (\cite{rellstab2020kids}),  do parental health shocks affect labor market outcomes of younger generation in developing countries? Moreover, do these effects disproportionately affect females? And to what extent do these impacts persist?  Empirical answers to these questions help elucidate the costs of family risk-sharing and offer insights into the potential dividends of market and social insurance reforms in the developing world.

To illuminate the behavioral responses of adult children's labor supply to parental health deterioration, this paper first constructs a conceptual framework grounded in the inter-temporal cooperative household model (\cite{chiappori2017static}). The framework indicates that, under the assumptions of: (1)intrahousehold cooperation; (2)differential market wage across gender; (3)existence of domestic production of eldercare, females are more inclined to exit the labor market as parental health declines. The framework allows for both a  substitution effect, which elevates the demand for informal care and curtails labor supply, and an income effect, which arises from heightened medical expenditures and motivates labor participation. These two effects bear significance not only for theoretical considerations but also for empirical and econometric practices. Empirically, existing research has predominantly scrutinized the substitution effect, specifically gauging the link between informal care and female labor supply. However, for a  comprehensive understanding of the overall repercussions of parental health shocks on adult children, considering the income effect is necessary. Methodologically, the coexistence of these effects presents challenges to conventional instrumental variable approach, rendering it unable to satisfy monotonicity and exclusion restriction assumptions. Consequently, resulting IV estimates are prone to bias and defy interpretation as local average treatment effects (LATE, \cite{imbens1994identification}).  Furthermore, the dynamic part of this framework considers the persistence of health shocks at older ages and endogenizes human capital accumulation of adult children, whereby parental health shocks tend to exert long-term negative influences on female labor supply under strong marital commitment.

This paper then leverages the staggered difference-in-differences design (hereinafter staggered DiD) to seek empirical causal evidence. The ideal data set would require detailed and longitudinal information on both parental health and children's economic and demographic variables. The longitudinal data set would also need to be mature with sufficient waves to capture any long-term effects and provide credible pre-trends tests.  Moreover, it must be nationally representative to shed light on the global consequence of population aging.  Our empirical analysis  rests on the China Family Panel Studies (CFPS) dataset spanning from 2012 to 2020, which uniquely meets these requirements. Following recent studies (\cite{rellstab2020kids},\cite{dobkin2018economic},\cite{garcia2013long}), we define parental health shocks as the initial hospitalisation of either adult children's parents or parents-in-law. This measure mitigates reported bias inherent in subjective health variables. We also demonstrate that hospitalization, while capturing severe symptoms, exhibits prevalence, persistence, and a significant upward trajectory with age.

 We investigate labor supply effects along both the extensive margin, gauged through employment status, and the intensive margin, measured by weekly work hours conditional on employment.  Our main results  reveal a noteworthy decline in the female employment rate, with an average reduction of 3.7 percentage points subsequent to parental health shocks. Conversely, there is no significant change in male adult children's employment. The results also find no significant changes in working hours. The shifts observed on the extensive margin are particularly likely to cause  scarring effects for future labor market performance. An important question is thus whether females affected by these shocks are capable of returning to employment quickly?  Additionally, any negative impacts can also be fueled by the persistent nature of health shocks at older ages.   We then undertake a dynamic analysis based on the event study approach.  Strikingly, our investigation demonstrates no evidence of recovery in female employment within the eight-year interval following a shock, the maximum span observable by our sample.

Subgroup analyses  unveil strong gradients in the employment effects by income and wealth. Notably, some subgroups turn out to increase their employment following a parental shock, indicating strong income effects.   Females with median income experience  a 4.77 percentage point decline in employment following a parental health shock, while for those with top decile income, the effects turn positive. The male labor income gradient remains absent. Intriguingly, male employment gradient emerges along household wealth. Particularly, males from households with assets below the median witness a 1.90 percentage point surge in employment post-shock.  Strikingly, this effect escalates to  6.02 percentage points for those from households with assets below the quintile threshold.  Meanwhile, no such gradients are found for females. These positive effects can hardly be explained by the rising demand for informal care, and are consistent with the existence of income effects posited by the conceptual framework. We also explore heterogeneous effects by education, marital status and  the distinction between parental and parents-in-law shocks.

In robustness analyses, this paper follows recent mythological advancements in DiD with treatment effect heterogeneity, i.e. \cite{callaway2021difference}, to rule out the ``forbidden comparison'', especially pertinent due to our keen focus on dynamic effects (\cite{sun2021estimating}). We also restrict the sample to individuals with no prior hospitalization records so that the shocks are more unanticipated. Additionally, the robustness analysis narrows the sample to adult children in prime working age, a cohort expected to demonstrate robust labor market attachment and resilience to adverse shocks.  Finally, we control for functions fully saturated in the age of each parent, a critical factor associates with both parental health and children's labor supply. The results remain robust across the various tests above.

This paper pioneers a causal exploration into the impacts of parental health shocks on the labor supply of adult children within the context of developing countries. While research has well noted the sensitivity of female labor supply to health shocks involving spouses (e.g. \cite{fadlon2021family}, \cite{coile2004health} ) or young children (e.g. \cite{breivik2022career}, \cite{eriksen2021impact}, \cite{gould2004decomposing}), this study reveals a new source of risk: parental health shocks, with particular concern under the rapidly aging populations.\footnote{It is noteworthy that \cite{wang2023consequences} find small and insignificant spill over effects of health shocks from the spouse in China. They define health shocks as the first onset of a cardiovascular or cerebrovascular health event.} Previous studies based on developed countries finds no effects or small gender disparity. Based on Netherlands, \cite{rellstab2020kids} find no significant effect of parental hospitalization on children's labor supply, attributed it to the country's well-established formal care system.  For Austria, \cite{frimmel2020health} find that both female and male's employment decline following a parental stroke, with modest gender difference.\footnote{They find a decline of two percentage points at 10\% significance level for female and 1.5 percentage points at 5\% level for male. However, they find a larger gap in earnings decline.} Interestingly, they find the decline disappeared after a liberalization of formal care markets.  The consequences of parental health shocks loom large and unequal  for developing countries, of which societies exhibit strong family ties and insufficient formal insurance.  Our finding, an persistent effect of approximately four percentage points that is specific to women,  highlighting the nontrivial and uneven burden under population aging.

More broadly, this paper adds to studies about adverse health shocks on labor supply, such as \cite{dobkin2018economic}  \cite{cai2014effects}, \cite{garcia2011institutions}, \cite{bound1999dynamic}, among others. These studies focus on the impacts on individuals themselves. Notably, using an event study approach, \cite{dobkin2018economic}  provide early evidences based on quasi-experimental design. They find  individual's  employment rate decreases by 8.9 percentage points one year after hospitalization, and 11.1 points three years later.

This paper also contributes to the literature about the effect of informal care on female labor supply, such as \cite{schmitz2017informal}, \cite{crespo2014caregiving}, \cite{van2013effect}. It's important to note that informal care, which constitutes the substitution effect within our framework, represents just one of the mechanisms through which parental health shocks take effects. We aim to assess the comprehensive influence of parental health shocks across both genders, with the income effect emerging as a pivotal determinant of gender difference. Additionally, this paper points out the potential econometric issues of ignoring the income effect. Research on informal care typically adopts parental health as an IV for care provision (e.g. \cite{crespo2014caregiving}, \cite{bolin2008your}), where the presence of income effects may invalidate the exclusion restriction and monotonicity assumption.

Another contribution of this paper is revealing the enduring impacts  through the lens of an event-study quasi-experimental design. Studies on the effects of informal care by IV typically estimate a static effect. The work by \cite{schmitz2017informal} is the only exception to our knowledge, which carefully selects control variables based on the conditional independence assumption. Our strategy relies on the parallel trends assumption and we provide credible tests up to eight years before the event. The implied costs of population aging may be far greater and uneven than the ones suggested by short-term effects, shedding light on the potential benefits that policies on long-term care insurance could yield.

Lastly, this paper offers a new  perspective to the literature on within-household labor division and gender gap, which attribute intrahousehold specialization to children, social normal, marriage quality etc. (e.g. \cite{juhn2017specialization}, \cite{bertrand2015gender}, \cite{yamaguchi2014labor}). This paper highlights the role of parental health via a theoretical framework combined with empirical  evidences.

The next section presents the theoretical framework. Section 3 describes the data and econometric methods employed. The empirical estimation results and robustness checks are presented in Section 4 and 5, followed by a summary in Section 6.

\section{Conceptual Framework}

	To gain insights for the impact of parental health on adult children's labor supply, we first construct an inter-temporal cooperative framework for household labor supply decisions (\cite{chiappori2017static}).  Marriage is prevalent in China and married couples accounts for 85\% observations of our sample. It also presents more intricacy in terms of decision making. The conceptual framework hence decides to focus on the work decisions of couples.  This paper focuses on any gender difference and long-term impact, thus intrahousehold specialization and dynamic decision-making are the key mechanisms to capture. To have a basic intuition, specialization arises from a combination of three elements: (1) the cooperative nature between spouses; (2) the presence of domestic production; (3) the differential wage distributions across genders. The key aspect of dynamic decision-making lies in that parental health shocks are persistent, and that future labor market conditions are endogenously determined by current decisions. Rational expectation is assumed. 
 
 We first lay out a static framework to focus on the impact of parental health on intrahousehold specialization and we extend this framework into dynamic to understand potential long-term implications.

	
	\subsection{Static Model}
		\subsubsection{Household Specialization}

The first element required for within-household labor division is the cooperation between husband and wife. It turns out that imposing the transfer utility property (TU) for the preference of husband and wife can facilitate the analysis without loss of generality. Under TU, the framework boils down to a unitary model, of which the household utility is derived from consumption, altruism, and leisure:
	\begin{align}
		U(c,w_i,w_j)=U(c)+\gamma U^p(H,d(w_i,w_j)) - \varepsilon_i w_i - \varepsilon_j w_j
	\end{align}

Where $ w_i,w_j\in\{0,1\}$, represent the labor supply of the husband and wife respectively. $\varepsilon_i$ and $\varepsilon_j$ capture the disutility associated with labor supply, reflecting idiosyncratic preferences for leisure. The consumption utility follows a generic concave function, characterized by $ dU/dc>0, d^2U/dc^2<0$. Parental utility enters the offspring household utility function in a caring form, with $\gamma$ representing the level of concern for the parental utility.\footnote{Caring form assumes that children only care about the utility of their parents instead of the decisions made by their parents. See \cite{chiappori2017static}.} It also reflects social norms in developing countries that are different from developed countries, such as filial piety.  Parental utility $U^p$ takes a parsimonious form to capture the utility of both parents and parents-in-law, which can be considered as a public good within the children household. Its original form is denoted as $U^p(H,d)$, depending on the health of parents H and the care provided by children $d$.  Parental health $H$ has a production function $H=h(z,d)$, where $z\in\{0,1\}$ indicates exogenous health status in period t ($z=1$ for poor health) and $d\in\{0,1\}$ indicates whether the parents receive caregiving from any children. If at least one spouse provides eldercare, it is considered that the parents receive care, hence we have $d=1-\mathbf{1}\{d_i=0\}\mathbf{1}\{d_j=0\}$, where $d_i,d_j\in\{0,1\}$ respectively represent whether each spouse provides informal care. Notice that the care provided by children may improve parental health or directly affect parental utility via accompany. For convenience, we redefine parental utility as a function of exogenous health status $z$ and informal care   $U^P(z,d) \equiv U^P(H,d)$. Regarding the function $U^P(z,d)$, we assume that it is supermodular and increasing in $d$, capturing the idea that the return from care is greater when parental health deteriorates:
	\begin{align*}
		U^p(z=1,d=1)- U^p(z=1,d=0) \ge   U^p(z=0,d=1)- U^p(z=0,d=0) \ge 0
	\end{align*}
The budget constraint of the household is given by:
	\begin{align}
	M(z)	+ c+\phi_i T^d_i d_i+\phi_j   T^d_j   d_j = \phi_i T_i+\phi_j T_j
	\end{align}
	  The wage rate for a unit time $\phi_i$, $\phi_j$ differs across genders because of different education level, work experience or discrimination. The second element required for labor division is that the wife faces a lower wage rate than the husband $\phi_j<\phi_i$. $T_i$ and $T_j$ are the total time endowments net of leisure that can be used for work or elder care, let $Y= \phi_i T_i+\phi_j T_j$. $T^d_i$, $T^d_j$ are the time needed for elder care, which are also allowed to be gender-specific. $M$ is the out-of-pocket medical expenditure, with $M(z=0)=0$ and $M(z=1)>0$.
	
	Now we consider the labor supply decisions of both spouses, focusing on the wife's decision for brevity. The model focuses on the binary choice between work and caregiving. Hence we have $d_j + w_j = 1$.\footnote{ We assume that it is difficult to simultaneously allocate time for both work and caregiving, with $T^d = T$, indicating that the total time available is fully devoted to either work or caregiving.} Firstly, when parental health is in good status and the husband works $w_i = 1$, the wife's employment decision is determined by comparing the utility under the following scenarios:
	\begin{align*}
		\text{\textit{Work:}}  \qquad U_{w_j=1 | w_i=1}=& U(Y)+\gamma U^p(z=0,d=0) - \varepsilon_i - \varepsilon_j \\
		\text{\textit{Care:}} \qquad   U_{w_j=0 | w_i=1}=&  U(Y-\phi_jT^d_j)+\gamma U^p(z=0,d=1)- \varepsilon_i 
	\end{align*}
When parents have good health $z=0$, the return to work for the wife, net of the disutility of work, is given by the following term:
		\begin{align*}
  S_{j}(z=0)\doteq\underbrace{U(Y)-U(Y-\phi_j T^d_j)  }_{\text{Gain in Consumption}} -  \underbrace{\gamma  [  U^p(z=0,d=1) - U^p(z=0,d=0) ] }_{\text{ Loss in Altruism}} 
	\end{align*}

When the wage rate of husband is higher than wife $\phi_i>\phi_j$ and the consumption utility function is monotonically increasing, intrahousehold specialization arises.\footnote{We assume the time spent for elder care is similar across gender $T_i^d=T_j^d$.}  As the return to work of the wife is smaller than husband $S_j<S_i$, the wife is more likely to exit the labor market. Consider the distribution of idiosyncratic disutility of work $\varepsilon_i,  \varepsilon_j \sim F(\varepsilon) $, the probability that a wife chooses to work given the employed husband is $	P(w_j=1| w_i=1)= \int \mathbf{1} (\varepsilon_j<S_{j} (z=0) ) d F_\varepsilon $.

\subsubsection{Poor Parental Health}

When parental health declines, how will couple's labor supply change? Firstly, the loss in altruistic utility resulting from work will increase due to the supermodularity property of $U^p$. When parents are sick, children are more likely to provide informal care and reduce their labor supply. This is the ``substitution effect'' focused by the literature on informal care and labor supply. However, at the same time, increased medical expenses resulting from poor parental health also reduce the overall household consumption, thereby raising the marginal utility of consumption and increasing the consumption gain from work. Consequently, parental health decline also generates an ``income effect''. Overall, it is indeterminate whether parental health decline reduces the overall return to work $S_{j}$, which depends on the quantitative significance between the gain in consumption and the loss in altruism:
		\begin{align*}
	\footnotesize S_{j}(z=1)\doteq\underbrace{U(Y-M(z=1))-U(Y-M(z=1)-\phi_j T^d_j) }_{\text{Gain in Consumption}} -  \underbrace{   \gamma  [  U^p(z=1,d=1) - U^p(z=1,d=0) ] }_{\text{Loss in Altruism}}
\end{align*}

Considering heterogeneity, individuals will exhibit different changes in their work decisions following a parental health shock. We first assume that parental health decline only causes a substitution effect, whereby $M(z=1)=0$. Notice that in this case, we have $ S_j(z=0)>S_j(z=1)$. Following \cite{angrist1996identification}, individuals can be categorized into three groups:
	\begin{itemize}
		 \setlength\itemsep{-0.1cm}
		\item Always-caregivers $\{  \varepsilon_j:  \varepsilon_j>S_j(z=0)   \} $ : Individuals that always provide eldercare regardless of parental health status
		\item Compliers  $\{  \varepsilon_j:  S_j(z=0)>\varepsilon_j>S_j(z=1)  \}  $ : Individuals that provide eldercare under poor parental health and work under good health
		\item Always-workers  $\{  \varepsilon_j:  \varepsilon_j<S_j(z=1) \}  $ : Individuals that always work regardless of parental health status
	\end{itemize}

\begin{figure}[H]
\centering
	\caption{Different Types of Children under Parental Health Decline}
			\begin{minipage}{0.75\textwidth}
	\begin{tikzpicture}[xscale=3]
		
		\draw(0,0)node[](x1){}
		--(0.95,0)node[](x2){}
		--(1,0)node[tick,label=above:\scriptsize{$S(z=1) $}](x3){}
		--(1.05,0)node[](x4){}
		--(2,0)node{}
		--(2.95,0)node[](x5){}
		--(3,0)node[tick,label=above:\scriptsize{$S(z=0) $ } ](x6){}
		--(3.05,0)node[](x7){}
		--(4,0)node[](x8){};
		\draw[decorate,decoration={brace,mirror,raise=5mm}](x1)--node[below,yshift=-6mm]{\text{\footnotesize Always-Workers}}(x2);
		\draw[decorate,decoration={brace,mirror,raise=5mm}](x4)--node[below,yshift=-6mm]{\text{\footnotesize Compliers}}(x5);
		\draw[decorate,decoration={brace,mirror,raise=5mm}](x7)--node[below,yshift=-6mm]{\text{\footnotesize Always-Caregivers}}(x8); 
		\draw[decorate,decoration={brace,mirror,raise=5mm}](x7)--node[below,yshift=-15mm]{\text{\footnotesize }}(x8);
	\end{tikzpicture} 
	\label{figure:complier}

 \tiny{This figure shows different types of children by their disutility of work. The horizontal axis characterizes individuals with different work disutility, from weak to strong and from left to right.}
			\end{minipage}
\end{figure}
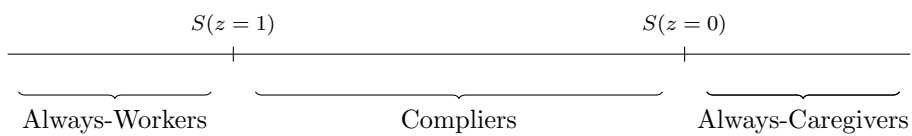

The categorization is represented by Figure \ref{figure:complier}. The horizontal axis represents different individuals with various levels of disutility of work $\varepsilon_j$, ranging from weak to strong from the left to the right. Regardless of parental health status, individuals with a high level of work disutility always choose to provide care, while those with a low level always choose to work. Under a deterioration in parental health, the work return decreases with $S(z=1)<S(z=0)$ , whereby individuals with moderate levels of work disutility reduce their labor supply.\footnote{For simplicity, we assume homogeneous return to work across individuals $S_j(z)=S(z)$. }

\subsubsection{Income Effect}

However, the deterioration of parental health also tends to inflict additional medical expense, resulting in a higher return to work. When the ``income effect'' is sufficiently large, there can be $S_j(z=1)>S_j(z=0)$. Some individuals may hence increase their labor supply after a parental health shock. This theoretical implication tends to have important implications for empirical analysis. For econometrics, using parental health as an IV of informal care in estimating its effect on labor supply is widely applied. The presence of this ``income effect'' invalidates the exclusion restriction and monotonicity assumption, making IV estimates biased and unable to be interpreted as a local average treatment effect (LATE).\footnote{It is a little bit nuanced whether the exclusion restriction or the monotonicity assumption will be violated. For instance, consider the estimation of the effect of hours spent on informal care on employment. If parental health only affects the intensive margin of labor supply but not the extensive margin, i.e. the employment, the exclusion restriction tends to hold but the monotonicity assumption is violated, as poor health may reduce hours spent on informal care following similar argument in our conceptual framework.  } For empirical analysis, individuals with lower household assets or higher wages are more likely to increase labor supply following a shock.  As this study focuses on impacts of parental health shocks for both genders, where there is a large wage gap, considering both the substitution and income effects is important for capturing the gender difference. 

\begin{figure}[H]
		\centering
						\caption{The Change of Work Probability after Parental Health Decline }
			\begin{minipage}{0.69\textwidth}
				\qquad \quad	\includegraphics[height=7.5cm]{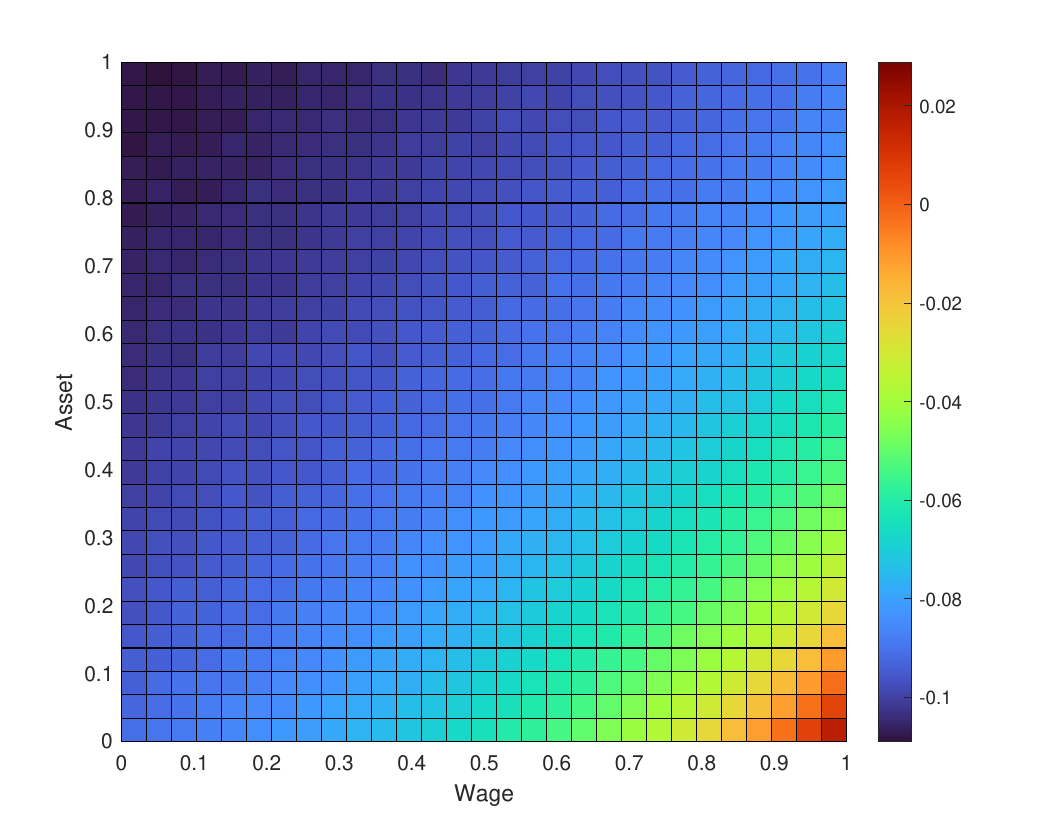}   
    
				\tiny{This figure shows how the   effect of parental health shock on work probability varies by wealth and wage. The result only aims to provide a qualitative insight. Household asset and individual wage are hence normalized between 0 and 1.  \par}	
			\end{minipage}
			\label{figure:defier}
	\end{figure}

Based on a simple structural model that is parameterized based on the theoretical framework, we simulate the changes in labor supply after a parental health shock. Figure \ref{figure:defier} confirms the heterogeneous employment effects of parental health deterioration. Individuals with lower household wealth and higher wages are less likely to leave the labor market, or even tend to increase their labor supply after a shock.

\subsection{Dynamic Model}
The second part of the theoretical framework extends the previous discussion into a dynamic setting, where all variables that vary over time are now  subscripted with t. Under inter-temporal decision making, the income in period t can be saved for future. Therefore, the budget constraint in Equation (2) is replaced by:
	\begin{align}
A_t	+ M_t(z)	+ c_t+\phi_{i,t} T^d_{i,t} d_{i,t}+\phi_{j,t}   T^d_{j,t}   d_{j,t} = \phi_{i,t} T_{i,t}+\phi_{j,t} T_{j,t} + (1+r)A_{t-1}
\end{align}
$A_{t-1}$ represents savings in the previous period, $r$ denotes the interest rate. One of the key mechanisms in this dynamic analysis is an endogenization of the wage rate $\phi_{j,t}$, allowing it to depend on experience. According to the classical Mincer equation (\cite{mincer1974schooling}), the wage rate is determined by education and work experience:
\begin{align}
	\phi_{j,t}=g(E_j, X_{j,t}), \quad  \text{ and } \quad  X_{j,t}=X_{j,t-1}+\delta w_{j,t} -  \lambda_t
\end{align}
$\lambda_t$ represents the rate of human capital depreciation, reflecting the natural changes in knowledge and skills over time. $\delta$ is the rate at which an individual's human capital accumulates when participating in work. If an individual remains out of the labor market for an extended period, their human capital will gradually decline, reducing the likelihood of re-entering the labor market in the future. Therefore, labor supply decision in the current period not only affects contemporaneous consumption and altruism utility but also has long-term implications by influencing the accumulation of an individual's human capital.

The second key dynamic feature is that the elderly health shock is more persistent than shocks at younger ages, as will be shown in the next section empirically. In addition, informal care can have a positive effect on the recovery of parental health.  Thus, the dynamic health transition exhibits the following properties:
\begin{align}
P(z_{t+1} =1 | z_{t}=1, d_t) &> P(z_{t+1} =1 | z_{t}=0, d_t)   \\
	P(z_{t+1}=1| z_t=1, d_t=1)& < P(z_{t+1}=1| z_t=1, d_t=0)
\end{align}
Considering the above dynamic mechanisms, value function of the household can be captured by the following Bellman equation:
	\begin{align}
	V_t(\Omega_t)= &	\operatorname*{max}_{c_t,w_{i,t}, w_{j,t} }  U(c_t)+\gamma U^p(z_t,d_t(w_{i,t}, w_{j,t})) - \varepsilon_{i,t} w_{i,t} - \varepsilon_{j,t} w_{j,t} \notag \\ 
 & \quad +  \rho   E  [   V_{t+1}( \Omega_{t+1}|  c_t,w_{i,t}, w_{j,t}, \Omega_{t}  )  ] 
\end{align}
$\Omega_t$ represents all the state variables in period $t$, including the parent's health status $z_t$, previous savings $A_{t-1}$, as well as the educational levels and work experiences of both spouses $E_i$, $E_j$, $X_{i,t}$, and $X_{j,t}$.\footnote{The savings transition equation is determined by the budget constraint equation (3), the work experience transition equation is determined by the endogenous human capital accumulation equation (4), and the parent's health status transition equation is determined by formula (5) and (6).} $\rho$ is the discount factor. Given the husband's employment, return to the wife's work decision is reflected by the following formula:
\begin{align*}
	S_{j, t} (z_t=1)\doteq & \underbrace{U(Y_t-M_t(z_t=1))-U(Y_t-M_t(z_t=1) -\phi_{j,t} T^d_{j,t} ) }_{\text{Gain in Consumption}} \\
	& \quad -  \underbrace{   \gamma  [  U^p(z_t=1,d_t=1) - U^p(z_t=1,d_t=0) ] }_{\text{Loss in Altruism}}  \\
& \quad	+  \underbrace{   \rho  E [  V_{t+1}( \Omega_{t+1}|  c_t,w_{j,t}=1, w_{i,t}=1, \Omega_{t} ) -  V_{t+1}( \Omega_{t+1}|  c_t,w_{j,t}=0, w_{i,t}=1, \Omega_{t} )  ] }_{\text{Return in Future}}
\end{align*}

In this dynamic framework, the wife's work decision $w_{j,t}$ will affect future returns through three mechanisms, leading to additional incentives compared to the static analysis. Firstly, it hinders parental health recovery, thereby also reducing future gain in altruism. This mechanism will encourage the children to reduce labor supply and increase eldercare. Secondly, although it increases current consumption, it is not conducive to reducing future medical expenses, which further makes the children more likely to reduce labor supply. Lastly, it contributes to the maintenance of human capital, thereby encouraging labor supply. However, it is noteworthy that while the first two channels have an impact on the entire household, the cost of the last channel is mainly gender-specific. While a decrease in the wife's human capital also has a negative effect on total household consumption, the first and second mechanisms tend to dominate especially if the wife has a low education. These dynamic channels will disproportionately generate long-term adverse impacts on employment of the wife.

\subsection{Features in Developing Countries}
In the end, this framework helps understand how the impacts of parental health shocks in developing countries can differ from developed countries. Notice that intrahousehold specialization relies on the degree of family cooperation, which in turn depends on the strength of marital commitment. In developing countries, due to different social norms and culture, as well as different economic foundations and legal systems, the level of marital commitment is often stronger. In addition, filial peity serves as a crucial social norm in China to encourage old-age support (\cite{guo2020effects}), suggesting a large weight of parental utility characterized by $\gamma$. The lack of social insurance and formal care also suggests $M(z)$ is larger in developing countries. At the same time, larger gender inequality leads to bigger gap in $\phi_j$ and $\phi_i$. All these factors combined,  parental health shocks are very likely to generate distinct implications in the current context.

 \section{Data and Econometric Method}
 \subsection{Data and Measures}

To quantify the impact of parental health risks on adult children's labor supply, the ideal data needs to satisfy three requirements. Firstly, it should provide detailed and longitudinal information on both parental health and children's labor market outcomes. Data sets that satisfy this requirement turn out to be rare. For instance, while China Health and Retirement Longitudinal Study (CHARLS) provides rich information on elderly health, information from children side is limited. As a result, this leads to the issue that only a few variables can be controlled for in any regressions at the children level.  For another example, China Labor-force Dynamics Survey(CLDS) provides detailed longitudinal information on labor market outcomes and children's demographics but the only health measure is the reported overall health. 
Secondly, as we are concerned with the consequence of population aging in a broad perspective, the data needs to be national representative to assess the overall impact. There are several sources of health insurance claim data in China, which provides administrative information on individual's specific health. However, these samples are typically based on population of a few developed cities and tend to miss the rural population, where family risk-sharing is more prevalent.  Finally, to capture any long-term effects as well as to provide credible tests on parallel trends, which is essential for DiD design, the longitudinal data set needs to accumulate a sufficient number of waves.

In this study, we utilize the China Family Panel Studies (CFPS)  data from 2012 to 2020. The CFPS is conducted by the Institute of Social Science Survey at Peking University and is a large-scale, nationally representative household panel survey covering various topics, such as economic activities, educational outcomes, family relationships, population migration, and health. The sample covers 25 provinces and includes data at the individual, household, and community levels. The survey design of CFPS has a particular focus on households, whereby detailed information for both parents and children is available.  While 25 instead of all provinces are surveyed, it covers 94.5\% of the population in mainland China, offering the chance to quantify the impacts of population aging of the whole country.  Most importantly, since conducted officially in 2010, this survey becomes mature and covers a span of ten years. These features uniquely satisfy the aforementioned research requirements.

In line with a number of studies,  this paper chooses the initial hospitalization of either parents or parents-in-law as the health shock event.  Initial hospitalization events have been widely used in recent research on the impacts of health shocks, such as \cite{rellstab2020kids}, \cite{dobkin2018economic} and \cite{garcia2013long}. As mentioned above, data sets that provide rich variables of children tend to have limited information on parental health. In studies about informal care and female labor supply, self-reported health is typically adopted. Hospitalization experience is plausibly more objective and less likely to suffer from reported bias. It also captures health shocks that tend to be more severe and persistent. For comparison, we also provide results based on the variable of self-perceived health decline in our later analysis. Due to data limitation, we cannot identify the specific symptoms that lead to hospitalization. Instead, we provide results based on health shocks preceded by years of absence of hospitalization records as robustness checks, where health shocks are more likely to be unanticipated.

\begin{figure}[H]
	\centering
	\caption{Trends of Hospitalization Risk and Likelihood of Informal Care}
	\begin{minipage}{0.99\textwidth}
		
		\includegraphics[height=6.2cm]{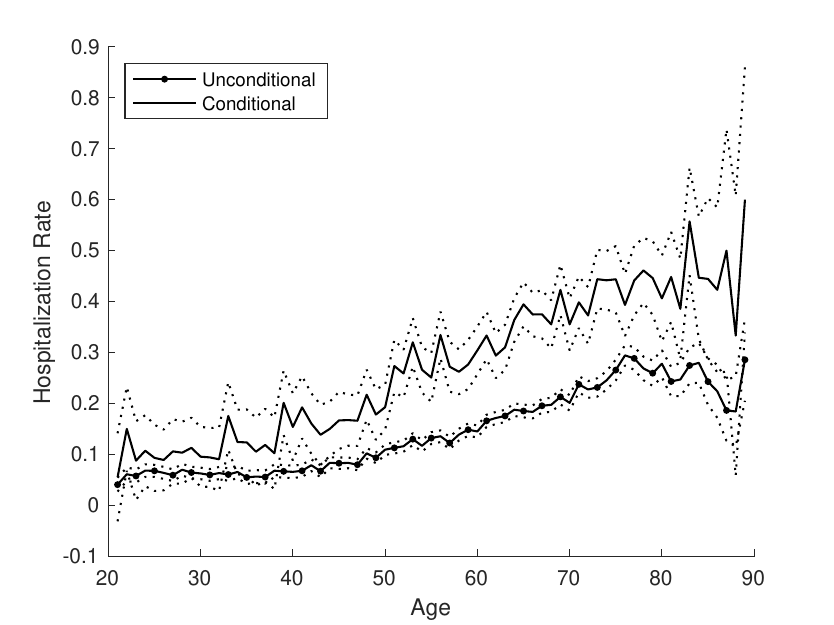} 	\includegraphics[height=6.2cm]{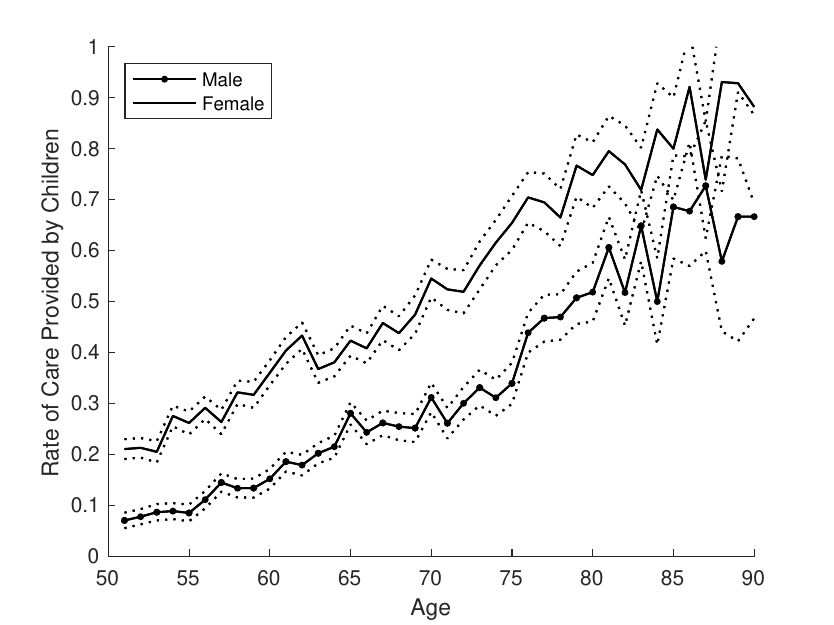}

		\tiny{The left figure shows the trends of hospitalization and the right figure shows the trends of informal care provided by children. Detailed definition about informal care variable is provided in Subsection 4.4.  Confidence intervals are at the 95\% confidence level. \par}	
	\end{minipage}
	\label{figure:facts}
\end{figure}

Hospitalization tends to reflect health shocks that are more severe, yet one may wonder about its frequency. After all, if it rarely occurs, the overall impact is inconsequential. The unconditional hospitalization rate  in Figure \ref{figure:facts} , i.e. the proportion of individuals that have experienced hospitalization during the past one year, shows a  clear rising trend over the life-cycle, reaching approximately 20\% at the age of 70.  More importantly, a special feature of parental health shock is that it tends to be more persistent than shocks happened at younger ages. The conditional hospitalization rate in Figure \ref{figure:facts} shows the hospitalization rate of the current period conditional on having inpatient experience in the last interview. The rate is significantly higher than the unconditional rate. Moreover, it also rises notably faster over the life cycle. As a result, parental hospitalization becomes not only more common but also increasingly persistent as individuals age. Coupling with the rising dependency ratio due to birth control policies such as the Later, Longer, Fewer in 1970s and the one-child policy in 1980s, the informal care burden induced by parental health shocks looms large.

While parental health shock is shown prevalent and persistent at old ages,  it has spillover effects if only children care about their parents and do take time to provide care. In Figure \ref{figure:facts}, the graph on the right shows that parents covered by informal care from their children account for a large and rising share. Approximately 50\% of mothers at the age of 70 receive care from their children. The result is calculated by fathers and mothers respectively. If we consider the union of parents and parents-in-law, the likelihood of caregiving for at least one parent of each children household tends to be even higher.

To construct our sample,  this study utilizes the data from 2012 to 2020 of CFPS. We drop the wave of 2010 due to different questions on labor market outcomes. Specifically, databases of the individual, family relationship, and economic conditions are merged for each year. Relevant information regarding health, labor market outcomes, and parent-child relationships for respective year is extracted from the merged data. Then, our sample uses each adult child as the basic unit of observation, matching with parental information. After obtaining the full sample, we restrict our final sample as adult children whose parents aged at least 50-year-old, with own ages between 16 and 64 years, and who have appeared in at least two survey waves. 
 The final dataset is an unbalanced panel dataset with 33,213 observations. Table \ref{table:vars} provides the explanation of each variable  used in following analysis, and Table \ref{table:des} provides the descriptive statistics.

\begin{table}[H]
	\centering
	\scriptsize
\caption{The Description of Main Variables}
	\begin{minipage}{0.99\textwidth}
		 \tabcolsep=0.12cm\begin{tabular}{p{0.25\textwidth}>{\centering\arraybackslash}p{0.70\textwidth}}
\hline
\hline
VARIABLES     &  DESCRIPTION    \\
\hline
Employment status & 1 for employed, 0 for unemployed or out of the labor force \\
Weekly working hours & Average number of hours worked per week \\
Age & Age of the respondent \\
Self-rated health & Self-rated health status of the respondent with five alternatives\\
Gender & 1 for male, 0 for female \\
Years of education & Number of years of education completed by the respondent \\
Marital status & 1 for married (having a spouse), 0 for unmarried, cohabiting, divorced, or widowed \\
Log of family assets & Natural logarithm  of the respondent's family assets\\
Number of children & Number of children of the respondent \\
Having children under 6 & 1 if the respondent has children below the age of 6, 0 otherwise \\
Urban/ Rural & Classification of residence as urban or rural \\
Number of siblings & Number of siblings of the respondent \\
Parental Health Decline & If respondents' parents or parents-in-law report that their health worse than a year ago \\
Parental Hospitalization  & If respondents' parents or parents-in-law had a hospitalization experience last year \\
Mother's age & Age of the respondent's mother \\
Father's age & Age of the respondent's father \\  
\hline
\end{tabular} \\
		\label{table:vars}
		\tiny{This table presents the main variables and their description.\par }
	\end{minipage}
\end{table}%

As Table \ref{table:des} shows, the average age of the respondents' father and mother is around 62 years. 58\% of the adult children show that their parents experienced a self-perceived health deterioration in the past year, while approximately 24\% show hospitalization in the same period.  Meanwhile, the proportion of individuals that have a child under 6 is 34\%. Approximately 53\% of the respondents were male, 47\% were female, with an average age of 35.6 years. At the time of the interview, 88\% of the respondents were employed, with an average of 48 working hours per week. Approximately 85\% of the respondents were married.

\begin{table}[H]
	\centering
	\scriptsize
\caption{Descriptive Statistics of Main Variables}
	\begin{minipage}{0.85\textwidth}
		 \tabcolsep=0.12cm\begin{tabular}{p{0.3\textwidth}>{\centering}p{0.12\textwidth}>{\centering}p{0.12\textwidth}>{\centering}p{0.12\textwidth}>{\centering}p{0.12\textwidth}>{\centering\arraybackslash}p{0.12\textwidth}}
\hline
\hline
Variable & Observations & Mean & Std & Min & Max \\
\hline

Employment status & 30,165 & 0.881 & 0.324 & 0 & 1 \\
Weekly working hours & 19,552 & 48.36 & 19.46 & 0.100 & 100 \\ 
Age & 30,165 & 35.61 & 8.740 & 16 & 64 \\
Self-rated health & 30,165 & 2.733 & 1.102 & 1 & 5 \\
Gender & 30,165 & 0.528 & 0.499 & 0 & 1 \\
Years of education & 30,165 & 9.450 & 4.273 & 0 & 23 \\
Marital status & 30,165 & 0.845 & 0.362 & 0 & 1 \\
Log of family assets & 30,165 & 12.59 & 1.38 & 0 & 17.75 \\
Having children under 6 & 30,165 & 0.338 & 0.473 & 0 & 1 \\
Number of children under 6 & 30,165   & 1.331  & 0.921  & 0 & 7   \\
Urban/ Rural & 30,165 & 0.488 & 0.500 & 0 & 1 \\
Parental health decline & 17,453 & 0.577 & 0.494 & 0 & 1 \\
Parental hospitalization & 17,451 & 0.239 & 0.427 & 0 & 1 \\
Father's age & 13,811 & 62.31 & 8.229 & 50 & 96 \\
Mother's age & 16,079 & 61.63 & 8.735 & 50 & 99 \\


\hline
\end{tabular} \\
		\label{table:des}

		\tiny{This table presents descriptive statistics of main variables for the estimation sample. Parental information is reported for own parents, as parents-in-law are simply own parents of the spouse.\par }
	\end{minipage}
\end{table}%

\subsection{Econometric Model}

\subsubsection{Baseline Model}
To analyze the impact of parental health shocks, our baseline econometric model is the Staggered Difference-in-Differences (Staggered DiD) approach. The econometric model is formulated as follows in a two-way fixed effect specification:
\begin{align}
    Y_{it} = \tau D_{it} + Z_{it}'\beta + \lambda_t + \alpha_i + \epsilon_{it} \quad 
\end{align}
In the equation, $Y_{it}$ represents the labor supply of child $i$ in period $t$, either the extensive or the intensive margin.  When the child's parents or parents-in-law have ever experienced a health shock, the variable $D_{it}$ is assigned the value 1. $Z_{it}$ represents a set of control variables related to the child, her family, and parents, such as age, marital status, education, logarithm of family assets, self-rated health, having children under 6 years old, urban-rural classification, and quadratic age function of each parent and parent-in-law. The individual fixed effects and time fixed effects are captured by $\alpha_i$ and $\lambda_t$, respectively. The parameter $\tau$ captures the average changes in labor supply for individuals that have gone through parental health shocks.

It is noteworthy that, instead of the average treatment effect(ATE), our quasi-experimental design can only identify the average treatment on the treated(ATT), which represents the average effect of parental health shocks on labor supply of children that had actually experienced the shocks. This is because the validity of our identification strategy relies on the conditional parallel trend assumption instead of a purely random assignment of treatment. For instance, parents with low socioeconomic status are be more likely to experience health shocks and their adult children may also have lower employment rates. While our identification strategy does not require that children of the treatment group have the same employment as the control group (only that they share similar trend in employment had the shocks not happen), the treatment and the control group may be nevertheless systematically different and so are the ATT and ATE. 

\subsubsection{Long-Term Effects}
In estimating long-term effects, this study employs the approach of Dynamic Staggered Difference-in-Differences, i.e. the Event Study (ES) method. The econometric model is presented as follows:
\begin{align}
Y_{it} = \sum_{p=k_1} \mu_p  D_{it}^{p} + \sum_{q=k_2} \tau_q D_{it}^{q} + Z_{it}'\beta + \lambda_{t} + \alpha_{i} + \epsilon_{it} 
\end{align}

Intuitively, the model compares the labor supply differences between the treatment and control group in each period against the benchmark period, which is set as two years prior to the initial hospitalization. 
$D_{it}^q = \mathbf{1}\left\{ t-E_i=q \right \} $ is an indicator function, where $E_i$ represents the period of event occurrence for the respondent $i$.  Similarly for $D_{it}^p$. The value of $D_{it}^q$ is equal to 1 when the respondent $i$ in period $t$ is $q$ years away from the event occurrence, where $k_1 \in \{ -4, -6, -8 \}$ and $k_2 \in \{0, 2, 4, 6, 8 \} $. Coefficients $\tau_q$ thus capture the dynamic effects of parental health shocks on adult children's labor supply and $\mu_p$ capture the pre-trends.

The series of coefficients $\mu_p$ can be used to test whether there is a significant difference in trends of labor supply between the treatment and control group prior to the occurrence of parental health shock. If a significant difference exists, the parallel trends assumption is most likely to be violated, and the trend of labor supply of the control group can hardly serve as a counterfactual reference for the treatment group after the shocks. If the trends are similar before the occurrence of parental health shock, it is plausible that these trends would continue had the shock not happen to the treatment group. If there are any concurrent changes in labor supply after the parental health shocks, the results are mostly likely to be driven by the health shock event.

\section{Empirical Results}
\subsection{Baseline Results}
Table \ref{table:sdid} presents results based on the baseline Staggered DiD regressions.  While our main focus is the extensive margin of labor supply, measured by individual's employment status, we also offer results about the intensive margin, which is the weekly working hours conditional on employment.\footnote{Different waves of CFPS provide different variables regarding working hours. We harmonize them into the weekly one, while a few outliers emerge. Weekly working hours variable are winsorized at 99 percentile to avoid these outliers.  }

 The results show that after experiencing a health shock to either their own parents or their parents-in-law, female adult children's employment rate decreases by an average of 3.73 percentage points, which is statistically significant at the 1\% level. To compare, this impact is as large in magnitude as having children under six years old, which has an effect of 3.92 percentage points. Meanwhile, due to intrahousehold specialization, marriage per se leads to a lower female employment rate by 13.27 percentage points, parental health shocks further reinforce this by 28\%. In contrast, we do not find  significant changes in male adult children's employment after a parental health shock. 

\begin{table}[H]
	\centering
	\scriptsize
\caption{The Baseline Results Estimated by Staggered DiD}
	\begin{minipage}{0.82\textwidth}
		 \tabcolsep=0.1cm\begin{tabular}{p{0.32\textwidth}>{\centering}p{0.15\textwidth}>{\centering}p{0.15\textwidth}>{\centering}p{0.15\textwidth}>{\centering\arraybackslash}p{0.15\textwidth}}
\hline
\hline

& \multicolumn{2}{c}{Working Hours} & \multicolumn{2}{c}{Employment} \\
& Male & Female & Male & Female \\
			\hline
			Parental Health Shock & -0.7743 & 0.3633 & 0.0034 & -0.0373*** \\
			& (0.7936) & (0.9910) & (0.0084) & (0.0143) \\
			Age & 0.0059 & 0.2649 & 0.0239*** & 0.0604*** \\
			& (0.6433) & (1.0803) & (0.0076) & (0.0127) \\
			Age Squared & 0.0014 & -0.0144 & -0.0003*** & -0.0009*** \\
			& (0.0065) & (0.0091) & (0.0001) & (0.0001) \\
			Marital Status & 1.1762 & -3.1059** & 0.0241* & -0.1327*** \\
			& (1.1887) & (1.5663) & (0.0142) & (0.0266) \\
			Years of Schooling & 0.4862 & 0.2431 & -0.0031 & 0.0068 \\
			& (0.4168) & (0.3743) & (0.0039) & (0.0060) \\
			Self-Rated Health & 0.3588 & 0.3991 & -0.0103*** & -0.0076* \\
			& (0.2715) & (0.3425) & (0.0029) & (0.0045) \\
			Ln Family Assets & -0.5618* & 0.0194 & 0.0014 & 0.0134*** \\
			& (0.3021) & (0.3576) & (0.0031) & (0.0050) \\
			Having Children Under 6 & -1.2050* & -0.8227 & 0.0055 & -0.0392*** \\
			& (0.6401) & (0.7855) & (0.0063) & (0.0121) \\
			Urban Hukou & 0.4205 & 1.8586 & 0.0190* & -0.0174 \\
			& (0.9894) & (1.3480) & (0.0102) & (0.0196) \\
			\hline
			Observations & 11,117 & 8,435 & 15,921 & 14,244 \\
			R-squared & 0.026 & 0.015 & 0.011 & 0.038 \\
			\hline
\end{tabular} \\
		\label{table:sdid}

		\tiny{This table presents the baseline estimates by Staggered DiD.  *** p$<$0.01, ** p$<$0.05, * p$<$0.1.  Second-order polynomial functions of the age of each parent or parent-in-law are also controlled for. Standard errors are clustered at the individual level. \par }
	\end{minipage}  
\end{table}%

We do not find statistically significant effects of parental health shocks on working hours conditional on employment. Literature on the effects of informal care typically finds less evidence with respect to the intensive margin, such as \cite{van2013effect} and \cite{ettner1996opportunity}.  The result is also consistent with the broad labor supply literature that usually finds smaller elasticity of intensive margin than extensive margin.

The results also reveal evidences of intrahousehold specialization in both intensive and extensive margin. While married females have lower employment rates than singles by 13.27 percentage points, married males' employment is 2.41 points higher than singles. In terms of intensive margin, married females who are employed work 3.1 hours less than singles per week, whereas married males work 1.2 hours more.

\subsection{Dynamic Effects and Parallel Trends}

\subsubsection{Long-Term Effects}

The baseline results reflect the changes in labor supply relative to the benchmark period averaged over all periods after a parental health shock. However, how long does the impact persist? In particular, after females became unemployed, how long does it take for them to return to the labor market? These long-term effects are particularly concerning. As shown in Figure \ref{figure:facts}, parental health shocks are more persistent than health shocks happened at younger ages, such as those to young children or adult spouse. This special feature tends to generate sustained impacts on females' employment. Further more, under intrahousehold specialization, which is prevalent among developing countries, informal care interrupts female's human capital accumulation. It leads to deteriorating labor market conditions and may further reinforce intrahousehold specialization, generating inconsequential and unequal impacts on females.

Table \ref{table:dsdid} presents the dynamic effects of parental health shocks estimated by the Event Study approach. The results show that female adult children experience a decline of 3.61 percentage points in employment rate in the year of the shock and 3.53 percentage points two years after. Although the estimates for four and six years after the shock are not statistically significant, mainly due to larger standard errors, there is no clear trend of recovery according to the magnitude. After eight years, the employment rate decreases by 6.94 percentage points and becomes statistically significant again.

Overall, there appears no evidence of a recovery trend in the employment rate of female adult children following a parental health shock  within the maximum observation period in our sample. These results indicate a sustained long-term impact of parental health shocks on female labor supply. In contrast, the results do not find a significant change in male employment rate following a shock. At the same time, there is no evidence of dynamic impact  on the working hours.

\begin{table}[H]
	\centering
	\scriptsize
	\caption{The Dynamic Effects Estimated by Event Study}
	\begin{minipage}{0.75\textwidth}
		\tabcolsep=0.15cm
		\begin{tabular}{p{0.25\textwidth}>{\centering}p{0.16\textwidth}>{\centering}p{0.16\textwidth}>{\centering}p{0.16\textwidth}>{\centering\arraybackslash}p{0.16\textwidth}}
			\hline
			\hline
		 &  \multicolumn{2}{c}{Working Hours} & \multicolumn{2}{c}{Employment} \\
			 & Male & Female & Male & Female \\
			\hline
			Before 8 years & -0.8641 & 2.8083 & -0.0219 & 0.0710 \\
			& (3.1223) & (3.4847) & (0.0335) & (0.0637) \\
			Before 6 Years & 1.0700 & 3.7196* & 0.0063 & 0.0066 \\
			& (1.7587) & (2.1442) & (0.0196) & (0.0318) \\
			Before 4 Years & -0.6369 & 1.5166 & -0.0168 & 0.0148 \\
			& (1.2499) & (1.5764) & (0.0132) & (0.0204) \\
			Before 2 Years & - & - & - & - \\
			& - & - & - & - \\
			Year of The Shock & -1.0235 & 0.9963 & -0.0033 & -0.0361** \\
			& (0.9360) & (1.1368) & (0.0096) & (0.0153) \\
			After 2 Years & -0.5085 & 0.3437 & -0.0017 & -0.0353* \\
			& (1.0883) & (1.3732) & (0.0110) & (0.0188) \\
			After 4 Years & -0.3900 & -0.6761 & 0.0069 & -0.0298 \\
			& (1.2956) & (1.6181) & (0.0132) & (0.0229) \\
			After 6 Years & -0.0474 & -0.2807 & -0.0131 & -0.0363 \\
			& (1.5548) & (1.9205) & (0.0172) & (0.0288) \\
			After 8 Years & -1.5236 & -1.3314 & -0.0184 & -0.0694* \\
			& (1.9515) & (2.5207) & (0.0210) & (0.0382) \\
			\hline
			Observations & 11,117 & 8,435 & 15,921 & 14,244 \\
			R-squared & 0.027 & 0.016 & 0.012 & 0.038 \\
			\hline
		\end{tabular} \\
		\label{table:dsdid}
		
		\tiny{This table presents the estimates of dynamic effects. *** p$<$0.01, ** p$<$0.05, * p$<$0.1. Control variables are the same as the baseline regression. Standard errors are clustered at the individual level. \par }
	\end{minipage}
\end{table}

\subsubsection{Parallel Trend Assumption}

	\begin{figure}[H]
		\centering
						\caption{The Dynamic Effects and Parallel Trends }
			\begin{minipage}{0.8\textwidth}
			
				\includegraphics[height=7.5cm]{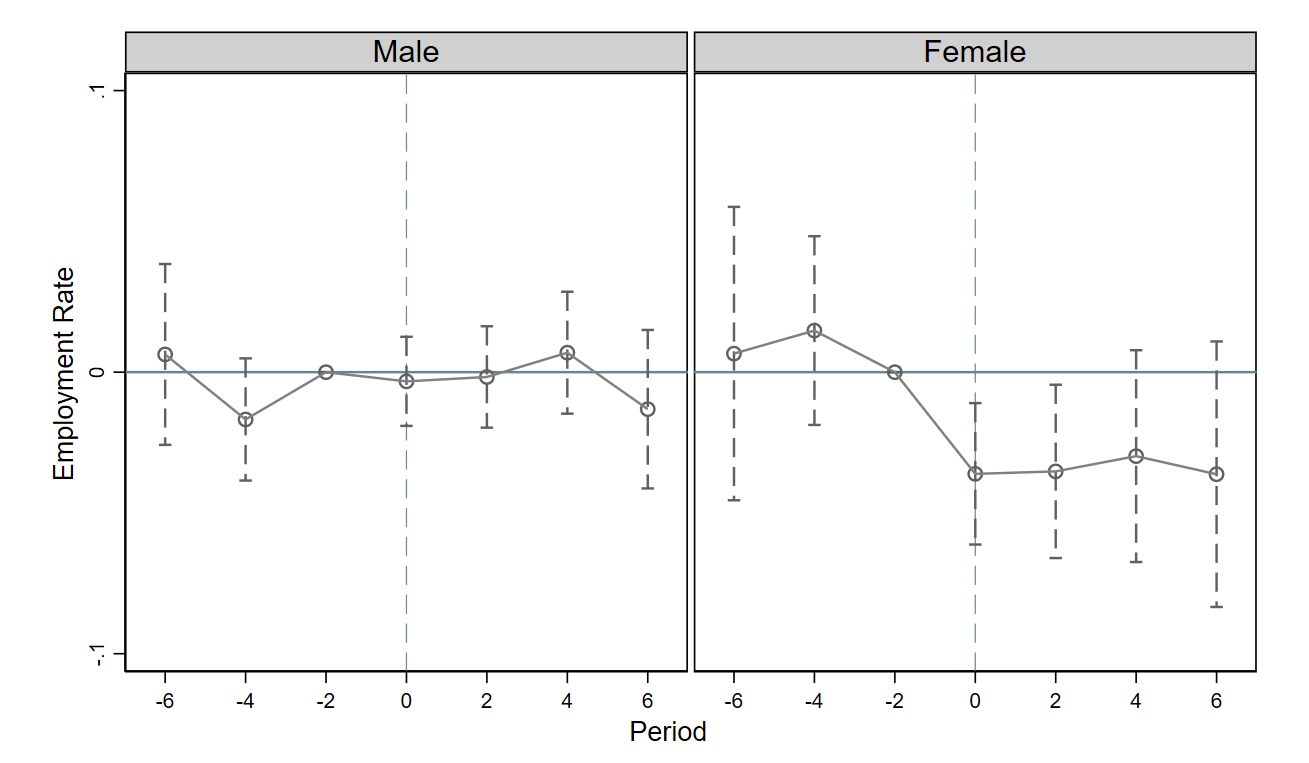}

				\tiny{This figure shows the trends of the difference in employment between the treatment and control group, for male and female adult children respectively. Confidence intervals are at the 90\% confidence level. \par}	
			\end{minipage}
			\label{figure:ptrends}
	\end{figure}

The core assumption of the difference-in-differences method is that the labor supply trend of the control group can be considered as the counterfactual trend for the treatment group if there were no treatment. Table \ref{table:dsdid} shows that, for both males and females, there is no significant differential trend in employment between the control and treatment group during the two to eight years prior to a parental health shock. It seems unlikely that the employment of the treatment group would experience a sudden and coincidental change for no reasons after the occurrence of a parental health shock. The most plausible explanation for these changes in employment should be the parental health shock. Figure \ref{figure:ptrends} provides a visual demonstration for the results of the event study estimation. It clearly shows a significant decline in female employment rate after the parental health shock, while there are no significant different pre-tends.

\subsection{Effects by Subgroups}
\subsubsection{Evidence of Income Effects}
As suggested by our theoretical framework, while one of the main implications of parental health shocks is the increased demand for informal care, which reduces work incentives, it may also lead to nontrivial medical expenditure, especially when health shocks are defined as hospitalization that captures severe symptoms. As a result, parental health shocks may not uniformly imply a lower employment rate due to the higher marginal utility of consumption. This is the key additional channel on top of the substitution effect focused by the literature on labor supply effects of informal care. To shed light on this mechanism empirically, we explore how does the employment effect of parental health shocks differ by individual income and by household wealth.

\begin{table}[H]
	\centering
	\scriptsize
	\caption{The Employment Effects by Individual Income and Household Wealth}
	\begin{minipage}{0.85\textwidth}
		\tabcolsep=0.02cm
		\begin{tabular}{p{0.38\textwidth}>{\centering}p{0.15\textwidth}>{\centering}p{0.15\textwidth}>{\centering}p{0.15\textwidth}>{\centering\arraybackslash}p{0.15\textwidth}}
			\hline
			\hline
		      	& Male & Female & Male & Female \\
			\hline
			\multicolumn{3}{l}{Individual Income}    &  &  \\
			\qquad Parental Health Shock     & 0.0012 & $ -0.0477^{***}$  & 0.0065 & $ -.0805^{***}$ \\
		                                       & (0.0092) & ( 0.0171) & (0.0153) & ( 0.0220) \\
			\qquad Shock $\times$ Log Income     & 0.0049 & $0.0399^{***}$    &         & \\
			                                      & (0.0101) & ( 0.0147)  &          & \\
			\qquad Shock $\times$ $\mathbf{1} \{ \text{Income}>\text{Median} \} $       & & &  -0.0089 & $ 0.0519^{*}$  \\
		   	& & & ( 0.0171) & (0.0290) \\
     \qquad  Mean Income, Yuan       & 44,452   & 29,820 & 44,452   & 29,820 \\
          \qquad  Median Income, Yuan      & 36,000   & 24,000 & 36,000   & 24,000 \\
             \qquad   Observations        &  12,802  &    9,697  &   12,802  &    9,697 \\            
			\hline
			\multicolumn{3}{l}{Household Assets}    & & \\
			\qquad Parental Health Shock      & -0.0085  &   $ -0.0388^{**}$ &    $-0.0231^{*}$     &    -0.0387     \\
		    	& (  0.0104 ) & ( 0.0189) &  (0.0140) &       (0.0285)  \\
			\qquad Shock $\times$ Log Asset          &  $-0.0346^{***}$   &   0.0006    &       &       \\
		    	& (  0.0133  ) & (  0.0230  )& & \\
			\qquad Shock $\times$ $\mathbf{1} \{ \text{Asset}<\text{Median} \} $     &    &  &  $0.0399^{**}$   &      -0.0005  \\
		   	&  & &    (0.0178)    &      (0.0328)  \\
             \qquad  Mean Assets, Yuan      & 263,510  &  263,510  &   263,510  &   263,510 \\
          \qquad  Median Assets, Yuan    & 239,080   & 239,080& 239,080  &  239,080 \\ 
                    \qquad Observations      & 44,452   & 29,820 & 44,452   & 29,820 \\
			\hline
		\end{tabular} \\
		\label{table:income}

		\tiny{This table presents how parental health shocks affect employment of individuals with different income and household assets.  For results by assets, we exclude individuals from the top quartile.  *** p$<$0.01, ** p$<$0.05, * p$<$0.1.  Control variables are the same as the baseline regression. Standard errors are clustered at the individual level.\par }
	\end{minipage}
\end{table}

The upper panel of Table \ref{table:income} shows the heterogeneous effects by individual income.\footnote{We construct the variable of log income as each individual's income under employment averaged over all periods, so that it is time-invariant .We also subtract the median of log income from it to facilitate the interpretation of the main effect.} For females that earn the median income,  we find the employment rate decreases by 4.77 percentage points, larger than our previous main estimates of 3.73 percentage points. However, for women with every one percent higher income, the likelihood of non-employment is reduced by 0.04 percentage points. For example,  for a female with annual income of 48,000 yuan, the decrease in employment rate following a parental health shock will be as small as -0.0077. Meanwhile, the employment rate decreases by as large as 8.05 percentage points for females with income lower than the median, and the effect for those with higher-than-median income is significantly smaller. Notably, these results imply a positive effect for females with income higher than 52,692 yuan.\footnote{The top decile of female income is 56,000 yuan in our estimation sample.}  While the income gradient is also qualitatively-consistent with a setting without income effects, as the heterogeneous effects can be driven by different opportunity costs of informal care, the positive employment effect cannot be explained with only the substitution effect.

The results by household wealth provide another piece of evidences about the income effects. The lower panel of Table \ref{table:income} shows that, while females from households with the median assets experience a 3.88 percentage points decline in employment following a shock, this effect does not differ significantly across households with different wealth level. Interestingly, while male employment does not change for median households, it increases by 0.0346 percentage points for households with every one percent less assets.\footnote{Because rich families tend to have different employment decisions, we exclude individuals with assets above the top quartile.}  In particular, the employment rate of males from households with lower-than-median assets increases by 1.68 percentage points following a parental health shock, in contrast to those with high-than-median assets of which employment possibility decreases by 2.31 percentage points.

\begin{figure}[H]
		\centering
						\caption{ The Income and Assets Gradients by Restricted Sample }
			\begin{minipage}{0.95\textwidth}
			
				\includegraphics[height=9.5cm]{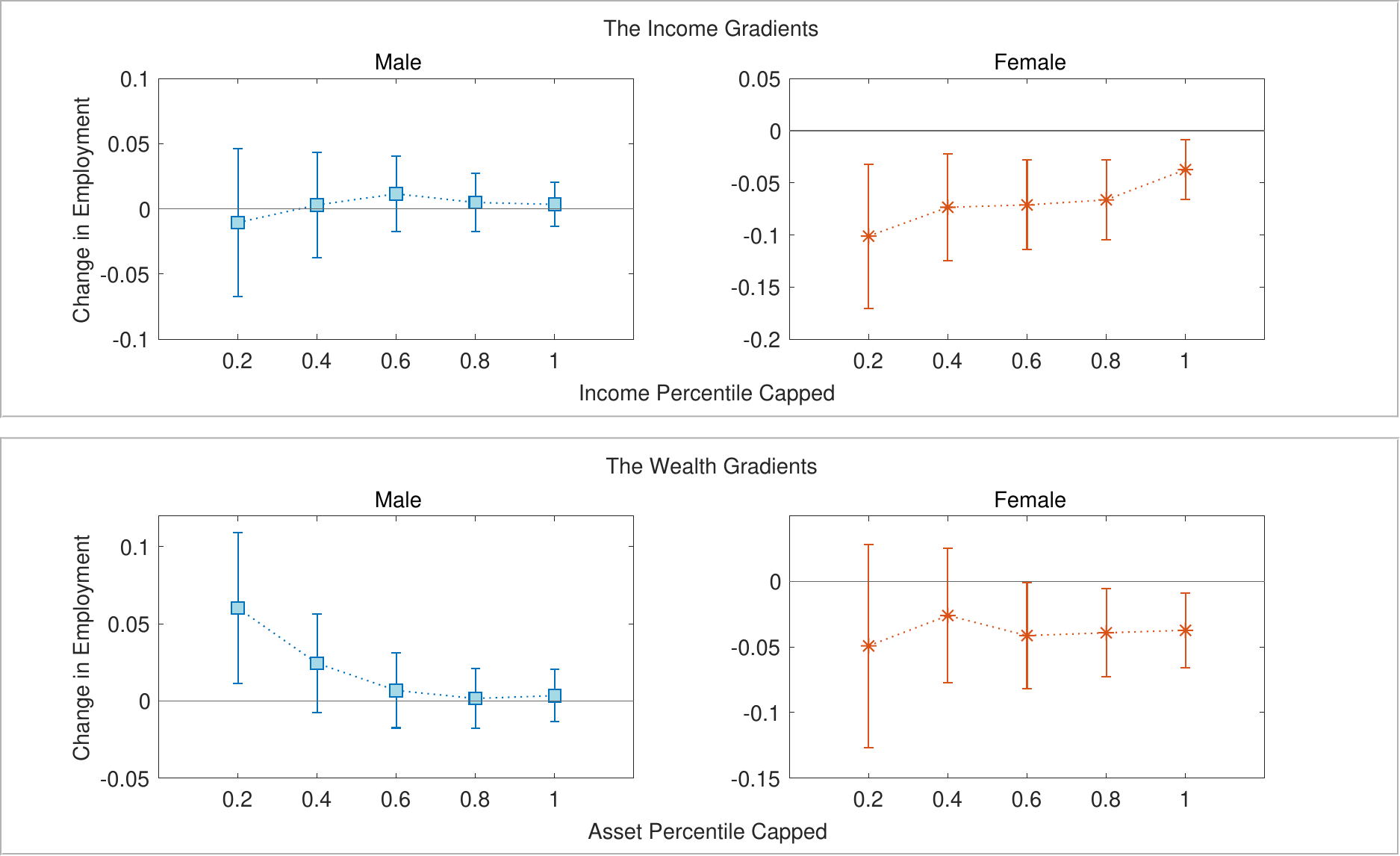}

				\tiny{This figure shows the income and wealth gradients in the employment effects of parental health shocks. Results are estimated on restricted sample with income or assets capped at corresponding percentiles. Confidence intervals are at the 95\% confidence level.\par}	
			\end{minipage}
			\label{figure:incomeeffects}
	\end{figure}

Figure \ref{figure:incomeeffects} provides a visualization of the income and wealth gradients, of which the results are estimated by excluding individuals with top percentile income or assets sequentially.\footnote{We provide results estimated by each quintile in the appendix. While the estimates are more noisy due to small sample size, the gradients are qualitatively the same. } As the sample is restricted to individuals with less income, the effects on male employment always remain close to zero. However, while the average impact on female employment is 3.73 percentage points, the one for females with income below the first quintile is as large as 10.11 percentage points.

The wealth gradients exhibit a different picture. The employment effects for female remain largely stable regardless of the sample restriction. Strikingly, while the employment of the whole sample of males does not change, we find those from families with less than the lower quintile of the assets distribution significantly increase their labor supply, with a hike of six percentage points. Overall, the positive employment effects of parental health shocks can hardly be explained by a higher demand for informal care alone. They are consistent with the existence of income effects and are important to capture the differential effects across genders.


\subsubsection{Other Heterogeneity}

Table \ref{table:hetero} further presents heterogeneous effects by education, marital status, and parent's type. Children with higher education levels tend to have higher labor income, and the opportunity cost of caregiving is higher, which should result in a milder reduction in employment after a parental health shock. Once considering the income effect, this negative impact will be further diminished or even reversed. In terms of marital status, intrahousehold specialization implies that the negative employment effect for married female may be larger. However, household economy of scale may also suggest more resources for risk-sharing, reducing the necessity for wives to exit the labor market. Finally,  if both spouses receive greater altruistic utility from their own parents and decision-making is non-cooperative, it is more likely that the health shocks of own parents produce a greater impact than those of parents-in-law.

\begin{table}[H]
	\centering
	\scriptsize
\caption{Employment Effects by Subgroups}
	\begin{minipage}{0.9\textwidth}
	   \tabcolsep=0.05cm\begin{tabular}{>{\centering}p{0.18\textwidth}>{\centering}p{0.15\textwidth}>{\centering}p{0.15\textwidth}|>{\centering}p{0.18\textwidth}>{\centering}p{0.15\textwidth}>{\centering\arraybackslash}p{0.15\textwidth}}
\hline
\hline
   & Male & Female &  &  Male & Female  \\
\hline
 \multirow{3}{*}{College} & 0.0141  &  -0.0263 &   \multirow{3}{*}{High School} & 0.0015  &  -0.0409***   \\
                          & (0.0183)	 &  (0.0306)  &            & (0.0095)	 &  (0.0162) \\
                        & Obs: 2,924      &   2,680     &              &  Obs: 12,997      &   11,564  \\
\hline
 \multirow{3}{*}{Married} & 0.0092  &  -0.0365** &   \multirow{3}{*}{Singles} & 0.0112  &  -0.0978*   \\
                         & (0.0090)	 &  (0.0151)  &            & (0.0267)	 &  (0.0500) \\
                         &Obs:  12,725     &12,777     &              & Obs:  3,196     &  1,467  \\
\hline

 \multirow{3}{*}{{Parents}} & 0.0025  &  -0.0316 &   \multirow{3}{*}{Parents-in-Law} & 0.0059  & -0.0445***   \\
                         & (0.0089)	 &  (0.0274)  &            & (0.0255)	 &  (0.0168) \\
                         &Obs:  14,262      &  4,273     &              &Obs:  1,796      &   10,112 \\
\hline
\end{tabular} \\
		\label{table:hetero}

		\tiny{This table presents heterogeneous effects by education, marital status and types of parental health shocks.    *** p$<$0.01, ** p$<$0.05, * p$<$0.1.  Control variables are the same as the baseline regression.  Standard errors are clustered at the individual level. \par }
	\end{minipage}
\end{table}%

The first row of Table \ref{table:hetero} shows that the employment of females with a high school degree or below are the most negatively affected, while males with a college degree show a positive effect, though the number is not precisely estimated. Again, this positive effect is difficult to be explained by the substitution effect alone. The second row of Table \ref{table:hetero} shows  no evidence that parental health shocks affect male employment, regardless of the marital status. In contrast, single females are more affected than married females. The estimates for single females are relatively imprecise, partly due to the smaller sample size. Finally, we do not find greater changes in employment in response to health shocks of own-parents, which suggests a high level of cooperation within the family.

\subsection{Elder Care}

The impact of informal care on female labor supply has received extensive attention in previous literature and is an important mechanism through which parental health shocks affect adult children's labor supply. We also explore the changes in informal care provision after the parental health shock. We exploit information from the question: ``Who took care of you when you were sick in the past year?" Possible answers to this question include: spouse, children or their spouses, grandchildren or their spouses, other family members, friends, social services, caregivers, parents, etc.   Based on this question, the dependent variable is generated to identify whether the children or their spouses provided care. Unfortunately, the survey does not differentiate whether the care is provided by the child or the spouse. To shed light on the gender difference, the following analysis thus focus on a sub-sample of unmarried children.

Table \ref{table:care} presents the estimates of the changes in probability of eldercare provided by unmarried children after  parental health shocks. It turns out that both  males and females significantly increase the likelihood of eldercare after the shocks, which suggest that the main reason for the decline in female's employment  may stem from their weaker labor market attachment. The caveat is that couples may have very different behaviors in terms of eldercare provision due to intrahousehold specialization. Therefore the results based on singles are provided as indicative facts and any extrapolation should be careful.

\begin{table}[H]
	\centering
	\scriptsize
\caption{The Effects on Eldercare Provision of Unmarried Children}
	\begin{minipage}{0.80\textwidth}
		 \tabcolsep=0.1cm\begin{tabular}{>{\centering}p{0.35\textwidth}|>{\centering}p{0.18\textwidth}>{\centering}p{0.2\textwidth}>{\centering\arraybackslash}p{0.2\textwidth}}
\hline
\hline
& & Male & Female \\
\hline
\multirow{3}{3.8cm}{Effects on Eldercare}   & Coefficients & 0.2435***  &	0.1675*** \\
   & & (0.0379)	& (0.0555) \\
& Observations & 3,476 &	1,716 \\
\hline
\multirow{3}{3.8cm}{Effects on Employment}   & Coefficients & 0.0112  &	-0.0978* \\
   & & (0.0267)	& (0.0500) \\
& Observations & 3,196 &	1,467 \\
\hline
\multirow{3}{3.8cm}{Implied Effects of Eldercare  on Employment}   & &   	&     \\
   &   Coefficients & 0.046  &	-0.584 \\
&   &   &	 \\
\hline
\end{tabular} \\
		\label{table:care}

		\tiny{This table presents results about the effect of parental health shocks on the likelihood of adult children's eldercare provision. Control variables are the same as the baseline regression.  Standard errors are clustered at the individual level. \par}
	\end{minipage}
\end{table}%

As we are able to estimate the effects of parental health shocks on both eldercare provision and employment, it is possible to estimate the effect of informal care on employment if one is willing to impose the monotonicity and exclusion restriction assumptions. The implied effects of informal care on employment are 0.046 for males and -0.584 for females.  To compare, \cite{crespo2014caregiving} find effects between -0.45 and -0.65 for females in southern European countries and \cite{carmichael2003opportunity} find an effect of -0.356 for the main female caregiver within the household but no significant gender difference. However, the result is significantly larger than the -0.04 by \cite{schmitz2017informal}. Once again, due to the income effect brought about by parental health shocks, the exclusion and monotonicity assumptions may not be satisfied. We nevertheless provide this result for comparison, as this IV estimate differs from previous ones in that it is based on the causal first-stage and causal reduced-form estimates obtained via DiD, rather than OLS by assuming random treatment.\footnote{More strictly, as IV approach assumes parental health as random, the estimates of the effects of parental health on caregiving and employment are the ATE, while our estimates by DiD are ATT. They are directly comparable only under homogeneous treatment effects.  Our identification assumption does not rely on random assignment of parental health shocks but instead on parallel trends, which is plausibly weaker.}

\section{Robustness}
\subsection{Treatment Effects Heterogeneity}

The identification of the main results primarily relies on the variation in labor supply of the treatment group relative to the control group before and after the health shock. In practice, the control group includes following cases: (1) observations of individuals who never experienced a parental health shock during the sample period; (2) observations prior to the parental health shock of individuals who experienced a shock in later period of the sample; (3) observations after the parental health shock of individuals who have already experienced a shock. \cite{goodman2021difference} notes that in the third case, if effects of the event are not homogeneous, such as that there are dynamic effects, estimation results of the static difference-in-differences can be biased. \cite{sun2021estimating} further point out that estimates of the dynamic difference-in-differences are also biased. In our context, it is highly unlikely that the labor supply changes of the adult children would remain constant after a parental health shock. Therefore even the parallel trend assumption holds, our previous results can be biased.

As a robustness check, we provide results based on the two-step approach proposed by \cite{callaway2021difference}, hereinafter as CS-DiD. The effects are firstly calculated based on each ``granular'' $2\times2$ DiD. Then they are aggregated. Specifically, this method removes the post-shock observations of individuals who have already experienced the shock from the control group. Furthermore, we select individuals who never experienced a parental health shock during the sample period as the control group, which yields cleaner and more interpretable estimation results.

\begin{table}[H]
	\centering
	\scriptsize
\caption{The Dynamic Effects Based on \cite{callaway2021difference}}
	\begin{minipage}{0.75\textwidth}
		 \tabcolsep=0.16cm\begin{tabular}{p{0.23\textwidth}>{\centering}p{0.16\textwidth}>{\centering}p{0.16\textwidth}>{\centering}p{0.16\textwidth}>{\centering\arraybackslash}p{0.16\textwidth}}
\hline
\hline
 &  \multicolumn{2}{c}{Working Hours} & \multicolumn{2}{c}{Employment} \\
 & Male & Female & Male & Female \\
\hline
			Before 6 Years  & - & -1.8239 & 0.0463 & -0.0792 \\
			& - & (5.6074) & (0.0386) & (0.0884) \\
			Before 4 years  & -2.3983 & -0.9773 & -0.0072 & 0.0017 \\
			& (2.9548) & (3.8554) & (0.0258) & (0.0406) \\
			Before 2 years  & 3.5283 & -2.5839 & 0.0104 & -0.0093 \\
			& (2.7132) & (3.2860) & (0.0167) & (0.0254) \\
			Year of The Shock & -1.9458 & 2.3140 & -0.0097 & -0.0381** \\
			& (1.4548) & (1.6309) & (0.0113) & (0.0174) \\
			After 2 years  & -1.8710 & -0.8657 & 0.0022 & -0.0472* \\
			& (1.9528) & (2.6355) & (0.0148) & (0.0249) \\
			After 4 years  & 2.2409 & -6.9264** & -0.0040 & -0.0432 \\
			& (2.2191) & (3.3901) & (0.0205) & (0.0353) \\
			After 6 years  & 0.3224 & 0.6186 & -0.0443 & -0.0502 \\
			& (3.4675) & (4.1383) & (0.0350) & (0.0523) \\
			observations & 5,226 & 3,730 & 10,063 & 9,186 \\
\hline
\end{tabular} \\
		\label{table:csdid}
		
		\tiny{ *** p$<$0.01, ** p$<$0.05, * p$<$0.1.  Control variables are the same as the baseline regression.  Standard errors are clustered at the individual level. \par }
	\end{minipage}
\end{table}%

\begin{figure}[H]
		\centering
						\caption{The Dynamic Effects Based on \cite{callaway2021difference} }
			\begin{minipage}{0.75\textwidth}
			
					\includegraphics[height=7cm]{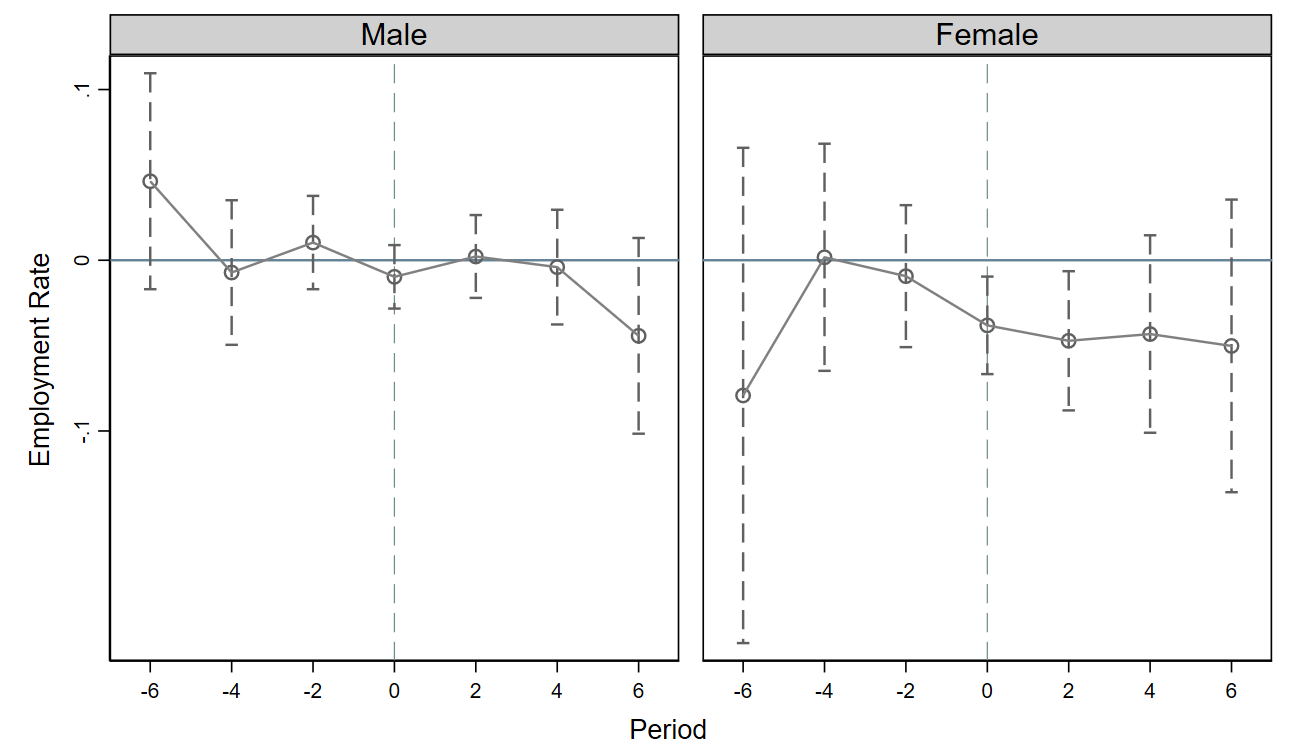}

				\tiny{ This figure shows the results estimated by the method proposed by \cite{callaway2021difference}. Confidence intervals are at the 90\% confidence level.}	
			\end{minipage}
			\label{figure:csdid}
	\end{figure}

Table \ref{table:csdid} presents the new results. Compared to females did not experience a parental health shock, the employment rate of the treatment group decreased by 3.81 percentage points in the year of the shock and by 4.72 percentage points two years after the shock.  Results about males show no significant changes. Although the use of a more stringent control group reduces the sample size and increases the standard errors of estimation results, the CS-DiD estimates are quite similar to the main estimates, demonstrating the robustness of the results. Figure \ref{figure:csdid} presents a visualization of the results. Again, we do not find significant difference in employment trends between the treatment and control group prior to parental health shocks.

\subsection{Unanticipated Hospitalization}

Our main analysis defines parental health shock as the initial hospitalization that is observed in the sample period. With a maximum of five waves of the data, it allows us to keep track of the dynamic changes in labor supply of an individual up to four waves following the parental health shock. It also enables us to test the trend of labor supply as many as four waves, approximately eight years, prior to the shock.

Despite this benefit, one may concern about whether adult children could anticipate the hospitalization of their parents. If they did, they may change their behaviors before the hospitalization, such as increasing their labor supply to save enough money for inpatient expenditure. While this concern is conceptually plausible, whether this bias exists and how much does it matter is an empirically question. If any, the changes in labor supply because of anticipation should invalidate the parallel trend assumption and fail our pre-trend tests.

\begin{table}[H]
	\centering
	\scriptsize
\caption{The Results Based on Restricted Sample with No Prior Hospitalization Records}
	\begin{minipage}{0.83\textwidth}
		 \tabcolsep=0.06cm\begin{tabular}{p{0.35\textwidth}>{\centering}p{0.15\textwidth}>{\centering}p{0.15\textwidth}>{\centering}p{0.15\textwidth}>{\centering\arraybackslash}p{0.15\textwidth}}
\hline
\hline
 & (1) & (2) & (3) &  (4)  \\
 \hline
  Male   &   &  &  & \\
\qquad	Parental health shocks& 0.0034 & 0.0015 &  0.0096 &  0.0079  \\
    &  (0.0084) &  (0.0089)  &  (0.0111)    &   (0.0158)  \\
\qquad			Observations & 15,921 & 12,051 & 9,621 & 8,250  \\
\hline
  Female   &   &  &  & \\

\qquad	Parental health shocks& $-0.0373^{***}$ &  $-0.0423^{***}$ &  $-0.0639^{***}$ &  $-0.0819^{**}$  \\
     &    (0.0143) & (0.0153)  &   (0.0206)  &  ( 0.0324)  \\
\qquad			Observations & 14,244 & 10,798 & 8,554 & 7,318  \\
\hline
No prior hospitalization records &   &   &   &   \\
\qquad At least 1 period (2 years) &   &  Yes &   &   \\
\qquad At least 2 periods (4 years) &   &   & Yes   &   \\
\qquad At least 3 periods (6 years) &   &   &   & Yes   \\
\hline
\end{tabular} \\
		\label{table:nohosp}

	\tiny{ *** p$<$0.01, ** p$<$0.05, * p$<$0.1.  Control variables are the same as the baseline regression.  Standard errors are clustered at the individual level. \par }
	\end{minipage}
\end{table}%

Empirically, we do not find systematically different trends in employment between the treatment and control groups, for both males and females. Indeed, while households may have leeway to some extent in deciding when to see doctors for some diseases, severe health issues that require inpatient treatments, such as stroke or heart disease, should be difficult to defer inter-temporally. Moreover, if any inter-temporal decision on hospitalization is possible, it is mostly likely to happen within months instead of across several years that our research design focuses. Due to the limitation of data, we could not identify the specific reasons for hospitalization. Nevertheless, we provide extra pieces of robust evidence by restricting the sample to parental health shocks that happened following years of absence of hospitalization, in which the anticipation of hospitalization is less likely.

Table \ref{table:nohosp} presents results based on the restricted sample that includes individuals with no hospitalization records at least for some periods prior to the initial one that we observe. The estimates are quite robust. If anything, the negative effects for females appear to be even larger as the sample restriction is more stringent. At the same time, the effects for men remain close to zero regardless of specifications.

\subsection{Prime Working Ages}
The main analysis restricts the age of children to be between 16 and 64 years old, aiming to quantify the global impacts of parental health shocks.  Younger individuals may have their labor supply influenced by going to school, while individuals aged 50 and above may be affected by retirement policies, as the statutory retirement age for female workers being 50, female officials being 55, and males being 60.  It is important to know whether our results are driven by these individuals who tend to have weak labor market attachment. Our previous findings are also more concerning if they hold among prime age workers. Therefore, this test further restricts the sample to individuals aged 25 to 50 years old.  Table \ref{table:primeage} shows that the average employment rate for prime age female workers decreases by 3.5 percentage points after the parental health shocks, which is statistically significant at the 5\% level and close to the main estimate of 3.73 percentage points.

\begin{table}[H]
	\centering
	\scriptsize
\caption{The Results Based on Prime Age Individuals }
	\begin{minipage}{0.90\textwidth}
		 \tabcolsep=0.1cm\begin{tabular}{p{0.3\textwidth}>{\centering}p{0.16\textwidth}>{\centering}p{0.16\textwidth}>{\centering}p{0.16\textwidth}>{\centering\arraybackslash}p{0.16\textwidth}}
\hline
\hline

& \multicolumn{2}{c}{Working Hours} & \multicolumn{2}{c}{Employment} \\
 & Male & Female & Male & Female \\
\hline
Prental Health Shock & -0.6635 &  0.0101 & 0.0575   & -0.0350** \\
& (0.8459) & (1.0514)  & (0.0149) &  (0.0086)  \\
\hline
Observations & 9,592 &  13,647 &  7,228  & 11,806 \\
R-squared & 0.024  & 0.008  & 0.013& 0.025 \\
\hline
\end{tabular} \\
		\label{table:primeage}

	\tiny{ *** p$<$0.01, ** p$<$0.05, * p$<$0.1.  Control variables are the same as the baseline regression.  Standard errors are clustered at the individual level. \par }
	\end{minipage}
\end{table}%

\subsection{More Rigorous Controls}

\begin{table}[H]
	\centering
	\scriptsize
\caption{The Results with Saturated Parental Ages and Schooling}
	\begin{minipage}{0.90\textwidth}
		 \tabcolsep=0.1cm\begin{tabular}{p{0.3\textwidth}>{\centering}p{0.16\textwidth}>{\centering}p{0.16\textwidth}>{\centering}p{0.16\textwidth}>{\centering\arraybackslash}p{0.16\textwidth}}
\hline
\hline

& \multicolumn{2}{c}{Working Hours} & \multicolumn{2}{c}{Employment} \\
Dependent Variable & Male  & Female  &Male  & Female  \\
\hline
Parental Health Shock & -0.5263 & 0.3828 & 0.0047  & -0.0377*** \\
& (0.8026) &  (1.0173) & (0.0085) & (0.0146) \\
\hline
Saturated Schooling & Yes & Yes & Yes & Yes \\
Saturated Parental Ages & Yes & Yes & Yes & Yes \\
Observations & 11,117 & 8,435 & 15,921 & 14,244 \\
R-squared & 0.057 & 0.066 & 0.035  & 0.058 \\
\hline
\end{tabular} \\
		\label{table:saturated}

	\tiny{ *** p$<$0.01, ** p$<$0.05, * p$<$0.1.  Control variables are the same as the baseline regression.  Standard errors are clustered at the individual level. \par }
	\end{minipage}
\end{table}%

In the main analysis, the age of each parent or parent-in-law is included in the regression in the quadratic form. The years of schooling of the children are controlled for as a continuous variable with linear slope. Both the age of parents and the education of children are critical variables in determining children's labor supply and may have nonlinear effects. This robustness test allows both of them to be fully saturated. Table \ref{table:saturated} shows that under stricter control of these variables, the employment rate of females decreases by 3.77 percentage points due to the parental health shock, which is significant at the 1\% level.

\subsection{Self-Reported Health Decline}

In this study, we define parental health shock as the initial hospitalization.  Self-rated health is a common measure of health and has been widely used in previous literature. In this subsection, we also explore results using the initial health deterioration reported by parents as the health shocks. Table \ref{table:slfh} presents results based on this definition.  It turns out that there is no evidence of a significant impact of this health shock even on female employment. While self-rated health may capture more comprehensive health changes compared to hospitalization, it may also include milder health changes. It is also likely to be subject to greater measurement errors. These results highlight the importance of using a more objective health measure and considering the consequences of more severe health shocks when assessing the impact of elderly health shocks.

\begin{table}[H]
	\centering
	\scriptsize
\caption{The Results Based on Self-Reported Health Decline}
	\begin{minipage}{0.90\textwidth}
		 \tabcolsep=0.1cm\begin{tabular}{p{0.3\textwidth}>{\centering}p{0.16\textwidth}>{\centering}p{0.16\textwidth}>{\centering}p{0.16\textwidth}>{\centering\arraybackslash}p{0.16\textwidth}}
\hline
\hline

& \multicolumn{2}{c}{Male} & \multicolumn{2}{c}{Female} \\
 & Working Hours & Employment & Working Hours & Employment \\
\hline
Parental Health Shock & 1.5975* & -0.0006 & 0.2572 & -0.0014 \\
& (0.8558) & (0.0091) & (1.0885) & (0.0150) \\
\hline
Self-Rated Deterioration & Yes & Yes & Yes & Yes \\
Observations & 9,592 & 13,647 & 7,228 & 11,806 \\
R-squared & 0.024 & 0.008 & 0.013 & 0.025 \\
\hline
\end{tabular} \\
		\label{table:slfh}

	\tiny{ *** p$<$0.01, ** p$<$0.05, * p$<$0.1.  Control variables are the same as the baseline regression.  Standard errors are clustered at the individual level. \par }
	\end{minipage}
\end{table}%

\section{Conclusion}

The proportion of aging population in China, along with many other developed and developing countries, is soaring.  The increasing incidence of elderly health shocks, the prevalence of risk-sharing via families, coupled with a rising dependency ratio, tend to generate spillover effects to other family members with large social costs.

This study provides causal evidences about the parental health penalty that is gender-specific and persistent. We first establish a conceptual framework based on inter-temporal household work decisions with cooperation, allowing for both "substitution" and "income" effects. It demonstrates that poor parental health may reinforce intrahousehold specialization, which may have long-term negative effects on female's labor supply. Using a unique data set which is recently mature for dynamic analysis, this paper examines the causal effects of parental health shocks on adult children's labor supply. Different from existing findings in developed countries, the empirical results reveal significant gender differences in employment changes following a parental health shock. Additionally, the dynamic analysis find no evidence of recovery up to eight years after the shock. We also find subgroups with rising employment rates following a shock, showing evidence of the income effect. These findings indicate the concerning consequence of ``growing old before getting rich'' for development countries and the necessity of improving social and market insurance systems.

\newpage

\appendix

\section{Income and Wealth Gradients}

	\begin{figure}[H]
		\centering
						\caption{ The Gradients in Income and Assets Estimated by Quntiles }
			\begin{minipage}{0.95\textwidth}
			
				\includegraphics[height=9.5cm]{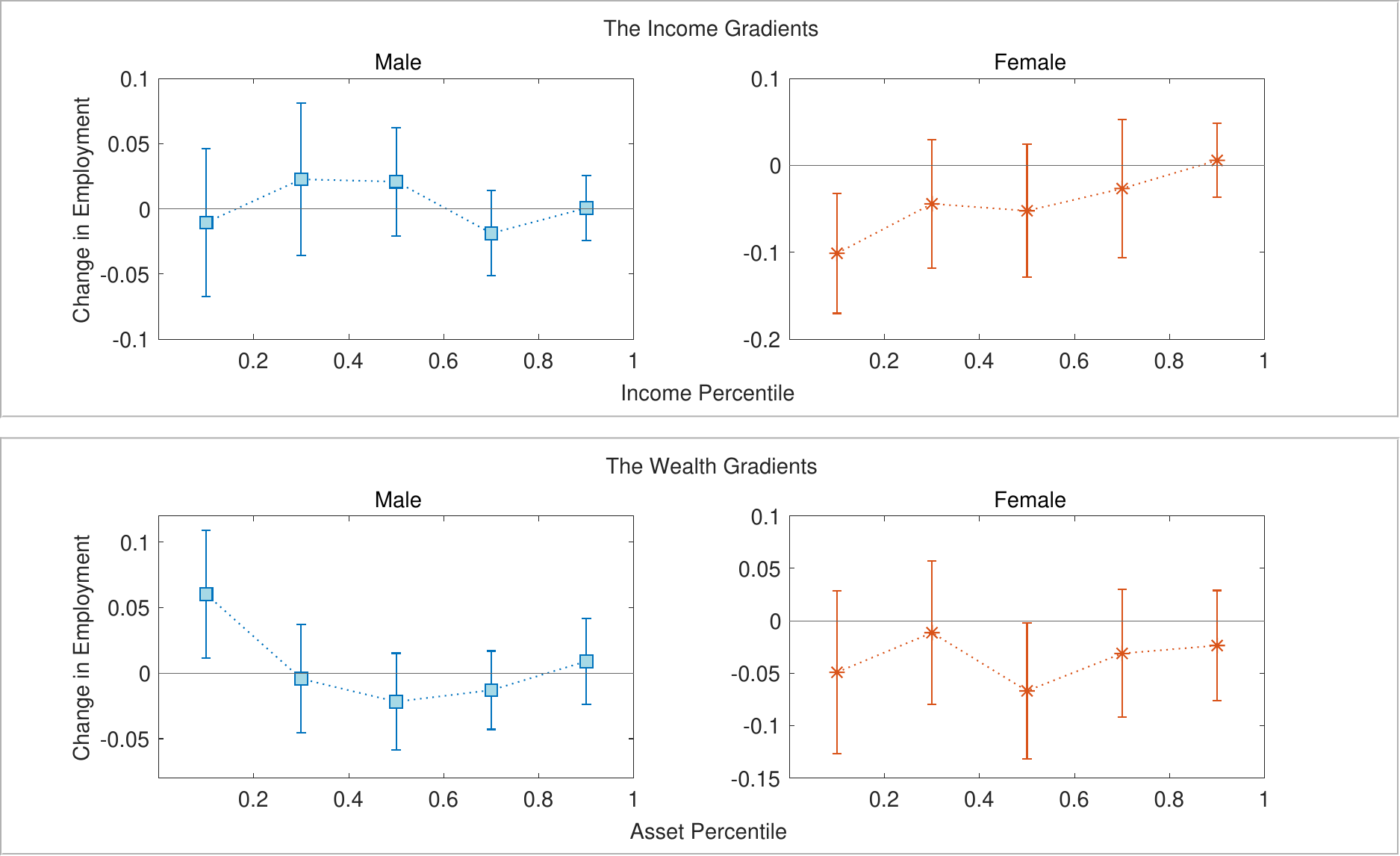}

				\tiny{This figure shows the income and wealth gradients in the employment effects of parental health shocks. The results are estimated by individuals from each quintiles. Confidence intervals are at the 95\% confidence level. \par}	
			\end{minipage}
			\label{figure:incomeeffect_B}
	\end{figure}

\newpage

\bibliographystyle{apalike}
\bibliography{Citation}

\end{document}